\RequirePackage{fix-cm}
\documentclass[twocolumn]{svjour3}      % twocolumn  %-------by me
\usepackage{graphicx}
\usepackage{subfigure}
\usepackage{xcolor}
\usepackage[switch, columnwise]{lineno}
\begin{document}
	
\title{Seasonal asymmetry in vertical distribution of meteor decay time at two conjugate polar latitudes}
\titlerunning{Seasonal asymmetry in vertical distribution of meteor decay time at two conjugate polar latitudes}   % if too long for running head

\author{Chenna Reddy Kammadhanam}
	\institute{Chenna Reddy Kammadhanam \\
		Department of Astronomy, Osmania University, Hyderabad, INDIA - 500 007.\\
		Tel.: +91-9010959080\\
		\email{chennaou@gmail.com}         
		}
	\date{Received: date / Accepted: date}
	\maketitle
	\begin{abstract}		
The meteor occurrence height and decay time height are strongly dependent on local atmospheric conditions in the mesosphere and lower thermosphere (MLT)-region. In this study, we comparatively examine the seasonal behaviour of vertical distribution of meteor occurrence height and decay time height at two identical radars of conjugate polar latitudes, Esrange (68$^\circ$N) and Rothera (68$^\circ$S). In order to understand the nature of meteor trail variations, the received signal power is categorised into two groups as weak and strong echoes, and their  seasonal mean vertical profiles are constructed. It has been noticed that the meteor occurrence height shown a seasonal symmetry, however, decay time vertical profiles shows an asymmetric pattern at conjugate polar latitudes, particularly for strong echoes. Seasonally, there is about 1 km difference in occurrence height and decay time height of weak and strong echoes. From the decay time vertical profiles, it has been noticed that the decay time turning altitude (i.e., inflection point) varies seasonally in the  altitudes range of 80-86 km for weak and strong echoes. 
%Seasonally, there is about 5 km difference in turning altitude across the latitudes, for both weak and strong echoes, and it gradually dropped down to 2 km. 
The maximum turning altitude of about 85 km is observed in Northern winter at Esrange (68$^\circ$N) and in Southern summer at Rothera (68$^\circ$S), similarly minimum turning altitude of about 80 km is observed in Northern winter at Esrange (68$^\circ$N) and in Southern summer at Rothera (68$^\circ$S).  %Above  the turning altitude, steeper slopes are observed for strong echoes, except in summer, at both the latitudes. %However, at Rothera, unlike the strong echoes, a large scatter is found for weak echoes below 70 km of altitude, except in winter season. 
The probable reasons for such behaviour of meteor trails at opposite polar latitudes are discussed.
		%\PACS{96.25.-f \and 96.30.Za \and Meteors, meteoroids, and meteor streams.}
\end{abstract}
	
\section{Introduction}
\label{intro}
It has been well established from the past several decades that the ablation of meteoroids in the mass range of 10$^{-11}$ to 10$^{-4}$g  are the main source of thousands of kilograms of meteoric mass flux per day in to the Earth's upper atmosphere (Plane, 2003). The dust particles of meteoric origin are responsible for a number of observed phenomena of the upper atmosphere and considered as the main source of neutral atom metallic layers distributed globally for the formation of sporadic-E layer (Haldoupis, 2011, Aylett et al., 2025).  Particularly, at high and polar latitudes, after ablation, meteoric dust particles undergo subsequent re-condensation into tiny particles of nanometer size and these condensed dust particles will act as the condensation nuclei for formation of ice particles (Rosinski and Snow, 1961, Berger and von Zahn, 2002). These charged ice particles are responsible for formation of noctilucent clouds (NLCs) and Polar Mesospheric Summer echoes (PMSEs) (Berger and von Zahn, 2002). For any given latitudes, properties of the meteoric mass flux is not uniform, as it originates from different radiant locations in the sky with specific orbital characteristics (Jones and Brown, 1993). The vertical size distribution of these dust particles is not uniform, which result in differential vertical distribution of meteor decay time. The meteor decay time governed by different physical processes at different altitudes, where ambipolar distribution dominates at lower altitudes and differential rotation dominates at higher altitudes (Sukara, 2013). Hence, it is important to understand the vertical distribution of the meteor decay time at two conjugate latitudes in order to understand and correlate with the atmospheric phenomena, to which it is responsible.\\

In recent past, all-sky interferometric meteor radars, developed based on the original antenna design proposed by Jones et al., (1998), become an important and potential tool to determine wide range of atmospheric parameters. A commercial version of these radars, known as SKiYMET radar, have been developed and deployed all around the world for the real time detection of meteor trail parameters. At present, more than 50 such radar systems are in operation to record meteor trail reflections in the height range between 70 and 110 km. These radars generally utilise almost isotropic wide beam transmitting antenna that pointed to zenith and capable of detecting meteor trails all over the sky by using five separate receiving antennas. Hence, these radars  could able to detect meteor echoes virtually in all directions of the sky. In addition, these radars record the radial drifts of individual meteor trails and are used to determine the large scale parameters of the mesosphere and lower thermosphere (MLT)-region. Over the years,  these radars proved to be very useful in survey of meteor shower radiant sources (e.g. Brown et al., 2008, Younger et al., 2009,  Campbell-Brown and Brown, 2014) and in determination of meteoroid orbital parameters (Janches et al., 2015), and also in detection of new showers  (Younger et al., 2015). By utilising ionisation height of shower meteors above sporadic background activity as a metric, Lukianova et al., (2018) and Reddy et al., (2019) has used data from these radars in recognition of established showers. All-sky radars have been employed successfully to study MLT-region winds (e.g., Kumar et al., 2007, John et al., 2011, Day et al., 2012, Korotyshkin et al., 2019, Ramesh et al., 2024), tides (e.g., Younger et al., 2002; Beldon et al., 2006, Liu et al, 2020), planetary waves (e.g., Pancheva et al., 2002; Lima et al., 2004, Kishore et al., 2018, Koushik et al., 2020) and gravity wave  variances and their momentum fluxes (e.g., Hocking, 2005; Beldon and Mitchell, 2009; Andrioli et al., 2015, Pramitha et al., 2019).\\

In particular,  meteor decay time of underdense echoes has been used in a number of studies.  To mention a few, underdense meteor decay time has been used to estimate atmospheric diffusion coefficient in the MLT-region (e.g. Chilson et al., 1996, Cervera and Reid, 2000, Ballinger et al., 2008, Younger et al., 2014). It is also used to determine the height of a mesospheric density level (Younger et al., 2015), in understanding the influence of geomagnetic storms on neutral mesospheric density at different latitudes (Yi et al., 2017, 2018a, b) and also to know the long-term variations in atmospheric density (Clemesha and Batista, 2006; Stober et al., 2012). Furthermore, by using a well established relation given by Mason and McDaniel (1988), diffusion coefficient values estimated from meteor decay time were used to calculate the neutral atmospheric temperature from known values of pressure or vice versa(e.g., Tsutsumi et al., 1994; Hocking et al., 1997; Cervera and Reid, 2000;  Hall et al., 2006, 2012; Holdsworth et al., 2006; Holmen et al., 2016) or to estimate the either pressure (p) or density ($\rho$) from known values of temperature (e.g., Hocking et al., 1997; Kumar, 2007; Takahashi et al., 2002). The same relation is used to estimate neutral density variations in mesosphere (Stober et al., 2012, Yi et al., 2018a, b) and also to determine the temperature-pressure parameter (Hocking et al., 1997; Kumar and Subrahmanyam, 2012). In this connection, it is worthy to mention the gradient method of Hocking (1999) to determine the temperature, which is independent of pressure values. \\ %Neilsen et al., 2001

However, all the above studies consider meteor decay time as a single entity, but the decay time profiles computed separately for weak and strong echoes based on received signal power have shown a distinct vertical distribution, particularly at mesopause region (Younger et al., 2008; Singer et al., 2008; Lee et al., 2013). Ballinger et al., (2008) computed decay time profiles for the weak and strong echoes at polar mesopause region and concluded that the processes other than the ambipolar diffusion can play a significant role in trail diffusion. From the systematic observations over two high latitudes (69$^\circ$N \& 67$^\circ$S) and a low latitude (22$^\circ$S) station, Singer et al., (2008) found a significant fall of decay time below 88 km of altitude for weak echoes than the strong echoes, at low latitudes throughout the year, and also at high latitudes with an exception of summer season. This differential behaviour of decay time is probably related to the presence of larger ice particles during the appearance of NLCs. Extending the similar study to two identical radars, one located at equator and the other at polar latitudes, Premkumar et al., (2019) found that the vertical profile turning altitude variation depends on seasons and found a difference of 2 km in the turning altitude of weak and strong echoes. In this context, the long term study of decay time vertical profiles at two identical radars of opposite latitudes is vey important. \\

Despite of deployment of identical meteor radars at opposite latitudes more than two decades,  most of the earlier studies were confined to understand the winds, waves and tidal activities in the MLT-region. There are no studies on direct comparison of decay time and the dynamical processes involved in decay time variation at opposite latitudes. In this context, here we studied the seasonal behaviour of meteor occurrence rate and vertical profiles of meteor decay time at Esrange (68$^\circ$N, 21$^\circ$E), Arctic and Rothera (68$^\circ$S, 68$^\circ$W), Antarctic. Because of identical design of radars, it is worth to compare the results directly to understand the different physical processes involved in meteor decay variation at opposite latitudes. In addition, a particular focus is on comparing and contrasting the meteor occurrence vertical profiles of weak and strong echoes. 
%----------------------- Radar System and Data Analysis-------------------------------
\section{Radar System and Data Analysis}
\label{sec:1}
The present study is based on data obtained from two identical all-sky SKiYMET meteor radars located at conjugate geographical latitudes, one at Esrange (68$^\circ$N, 21$^\circ$E), Arctic and the other at Rothera (68$^\circ$S, 68$^\circ$W), Antarctic. Both the radars  operate at a frequency of 32.5 MHz with a peak transmitting power of 6 kW having  a duty cycle of 15\%, and uses a transmitted radio pulse of PRF (pulse repetition frequency) of 2144 Hz with a pulse length of 13.3 $\mu$s that gives a range resolution of 2 km. The antenna system consist of a crossed-dipole Yagi transmitting antenna (Tx) of 3-elements each and five separate crossed-dipole Yagi receiving antennas (Rx) of 2-elements each. The Tx-antenna has a wide-beam with all-sky illumination pattern to scan large portion of the sky. The Rx-antennas form an interferometric array from which the unambiguous angle-of-arrival can be determined from phase difference of the received signal between each antenna pairs (Jones et al., 1998). By using a careful selection criterion designed in data acquisition software, the radar system exclusively records only underdense meteor echoes with echo duration less than 2 s. Further technical details and meteor detection procedure with SKiYMET radar can be found in Hocking et al., (2001).\\

The archival data were obtained on special request. Both the radars have produced relatively continuous data for about 15 years during 2005 - 2020. Despite of similar design, the meteor detection rates were quite different at both the radars. The Esrange radar detected about 12 000 -18 000 echoes per day whereas at Rothera it is about 10 000 - 20 000 echoes per day, however, the flux variation has similar diurnal and seasonal pattern.  To increase the data reliability and to avoid uncertainties in meteor detection heights, for further analysis, we have selected only unambiguous echoes with zenith angle in the range of 10$^\circ$ - 60$^\circ$. Furthermore, the total data were divided into two groups as weak and strong echoes, based on backscattered received signal power for every one kilometre height bin. The strong echoes are those with highest 25\% of received echo signal power, and the weak echoes are those with lowest 25\% of received echo signal power. A simple classification of meteor echoes into weak and strong categories based on the upper and lower quartiles of received signal power may introduce seasonal or height-dependent biases. These biases arise from the variations in relative signal power due to changing atmospheric conditions and background variability. To minimise such biases, the classification scheme was applied separately for each season and for every one-kilometre height bin. However this is not an absolute categorisation, but  a simple method of dividing the total number of echoes into approximately equal groups (Ballinger et al., 2008). This approach of categorisation of decay time as weak and strong provides a statistically balanced framework for comparative analysis. A similar categorisation of meteor decay time was adopted by Younger et al., (2008), Kim et al., (2010) and Premkumar et al., (2019). 
%-------------------------------------------Results and Discussion------------------------
\section{Results and Discussion}
\label{sec:2}
A long term study on seasonal distribution of meteor count rate at identical radars of opposite latitudes can be useful to compare and contrast the performance of the radar systems. Figure 1 shows a 15 years composite daily mean meteor count at Esrange (68$^\circ$N, Fig. 1a) and Rothera (68$^\circ$S, Fig. 1b) during 2005 - 2019. The meteor count show a clear seasonal pattern across the latitudes. We can also notice significant short term alterations and annual variation of meteor count rate, with minimum count rate coincides with the vernal equinox at both the latitudes. On the other hand, maximum is close to summer solstice at Esrange and it is about one month after the summer solstice (January) at Rothera. In addition, the centre of the broad peak (high plateau) is around autumnal equinox at both the hemispheres. Besides the seasonal pattern, meteor count also shows large annual variation with maximum to minimum count rate ratio of 2.5 and 2.2, respectively at Esrange and Rothera. The seasonal pattern of count rate is a result of  combined effect of dissimilar distribution of sporadic radiant sources at given latitude (Szasz et al., 2005), annual changes in atmospheric density and sensitivity of the radar systems (Younger et al., 2009). The annual meteor count rate at Esrange (68$^\circ$N) shows an increasing trend until June and then it is continue to decrease to the end of the year. In contrast, the count rate  at Rothera (68$^\circ$S) shows a decreasing trend till September and then it start increasing to the end of the year. Except the sudden shower peaks, the count rate at both the latitudes mainly  seasonal dependent.\\

As mentioned earlier, at both the latitudes, there are sudden intense peaks of high count rate due to major meteor showers, which are easily identified on the annual sporadic background activity. At first glance, the annual activity of some of the major showers can be easily noticed as strong peaks in figure 1. The most intense peaks identified in 15 years (2005 - 2019) mean meteor count rate are due to the Quadrantids meteor shower that occurs in early January and the Geminids in mid-December at Esrange (68$^\circ$N) as shown in figure 1c. Similarly, an intense peak due to South $\delta$-Aquarids meteor shower that occur in late July at Rothera (68$^\circ$S) as shown in figure 1d. There are other established but less intensive peaks are also noticed. For further detailed identification of such shower peaks, the trail ionisation heights are used as metric, where some of less active shower peaks are clearly manifested on sporadic background activity, as the shower meteors ionise at higher altitudes than the sporadic meteors (Hawkes and Jones, 1975, Lukianova et al., 2018).\\

Using the median heights of meteor trail ionisation and their corresponding upper and lower quartiles, Lukianova et al., (2018) recognised all major meteor showers of northern hemisphere from Sondankyla Geophysical Observatory (SGO, 67.4$^\circ$N,  26.6$^\circ$E) meteor radar. Similarly, Reddy et al., (2019) comparatively examined the trail ionisation heights and identified major showers at high (Eureka, 80$^\circ$N) and low latitudes (Thumba, 8.5$^\circ$N) of northern hemisphere. All these studies indicates that the higher ionisation heights of shower meteors are likely due to higher speed and porous bodies of high fragile nature of the shower meteoroids (Hawkes and Jones, 1975). In addition, ionisation heights also depend on the atmospheric density, meteoroid composition, radar transmitting power and frequency,  and also depends on meteoroid properties (Stober et al. 2014). In order to identify less active showers on background sporadic activity, we have calculated the median (M) and corresponding lower (Lq) and upper quartiles (Uq) of the meteor ionisation. Where,  M varies in the range of 87-91 km, while Lq and Uq vary in the range of 82-89 and 91-99 km, respectively.  Figure 2 shows a 15-years composite median (M), Lq and Uq of meteor trail ionization heights and corresponding residual plots determined from the superposition of daily meteor count at Esrange (68$^\circ$N, Fig. 2a) and Rothera (68$^\circ$S, Fig. 2b).  The composite plots computed so reveals the regular annual features by smoothing out inter-annual variations.\\ 

From figure 2, it can be noticed that there is a wide scatter in trail ionisation heights with large annual variations at both the latitudes. The maximum to minimum variation in median (M) height is about 2 km, whereas it is about 2-3 km in Uq and Lq. Within the annual variation, comparatively small and narrow peaks of one to several days are identified. These peaks are coinciding with the days of known showers, where major showers are identified and marked in Fig 2c-2e and Fig 2f-2h, their corresponding shower dates, taken from the literature, are given in Table 1. About eight showers are identified at Esrange (68$^\circ$N) and five showers at Rothera (68$^\circ$S). Most of these showers ionise in Uq, with an exception of the $\gamma$-Ursae minorids and Draconids ionise in median (M) height and $\sigma$-Hydrids ionise in Lq. A very few showers are identified to ionise in Lq height. All the marked showers at Rothera ionise at Uq. As marked  in figure, three unknown peaks are identified, one at Esrange and two at Rothera, where the days of ionisation peaks  are not coinciding with any known annual meteor showers.\\ 

In order to understand the seasonal variation of vertical distribution of meteor trails, the meteor trail occurrence height profiles were computed separately for weak and strong  echoes. As outlined earlier,  based on received signal power, the total data were divided into two groups as weak and strong echoes. A 15 years representative seasonal mean plots are as shown in figure 3, where the distributions are normalised to their peak value in each 2 km height bin and error bars represents standard deviation values. Here, winter (November - January), spring (February - April), summer (May - July), fall (August - October) are the seasons at northern hemisphere (Esrange, 68$^\circ$N). Similarly, winter (May - July), spring (August - October), summer (November - January), fall (February - April) are the seasons at southern hemisphere (Rothera, 68$^\circ$S). As both the stations are well in Arctic/Antarctic circle, which experiences only two distinct seasons, summer and winter, however for our practical purpose, we have chosen four seasons so that the equinoxes and the solstices are in middle of three-months window.  From  the figure 3, we can observe a clear seasonal variation in peak occurrence height for weak and strong echoes. At Esrange (68$^\circ$N), it is about 88.5 - 89.0 km for weak echoes and it is between 90.5 - 91.0 km for strong echoes. There is about 0.5 - 1 km  difference between the peak occurrence height of weak and strong echoes, which is  seasonal dependent, the difference is minimum in Northern winter, spring and summer (about 0.5 km) and maximum in Northern fall (about 1 km). A similar seasonal difference of 0.5 - 1 km is also found at Rothera (68$^\circ$S), but the peak occurrence height is few hundred meters higher, particularly in Southern summer. From figure 3, a difference  of few kilo-meters between the peak occurrence height of weak and strong echoes is evident, irrespective of latitudes. It can be inferred that there is no latitudinal difference, but only a seasonal difference in peak occurrence height of weak and strong echoes. The symmetrical distribution of peak occurrence heights at conjugate polar latitudes may be attributed to the comparable transmitting power and configuration of the radar sytems. Because of seasonal density variations, Younger et al., (2008) reported a mean detection altitude of about 3 km lower in Southern winter than Southern summer at Davis station (68$^\circ$S).\\ 

In addition, to understand seasonal behaviour of meteor decay time, received signal power vertical profiles are constructed separately for weak and strong echoes. Again, a 15 years representative seasonal mean meteor decay time vertical profile plots for both the latitudes are as shown in figure 4. Here, the decay time data were averaged in each 2 km height bin so as each data point indicates the seasonal mean and error bars represents standard deviation, similar as earlier. At Esrange (68$^\circ$N), it is noticed that the turning altitude (inflection point), a point from where meteor decay time start to decline, varies seasonally between 79 - 85 km for both weak and strong echoes. Precisely, the turning altitude for weak echoes varies between 79 - 84 km, with maximum turning altitude in Northern winter and minimum in Northern summer. For strong echoes, it is varies between 81 - 85 km having similar seasonal pattern as weak echoes. Similarly, at Rothera (68$^\circ$N), the turning altitude varies seasonally between 80 - 86 km for both weak and strong echoes, with maximum in Southern summer and minimum in Southern winter. Precisely, for weak echoes it varies between 80 - 85 km and for strong echoes it varies between 81 - 86 km. At both the latitudes, a difference of about 1 km can be noticed between weak and strong echo turning altitudes across the seasons. It is observed that the turning altitudes are having seasonal asymmetric pattern at opposite polar latitudes. However, there is about 1 km difference in turning altitudes of both weak and strong echoes irrespective of the seasons, but the turning altitudes at Rothera (68$^\circ$S) are few kilo-meters higher than the same at Esrange (68$^\circ$N).\\

A large scattering in meteor decay time is observed, particularly below 75 km and above 95 km of altitude, where decay time profiles exhibit high uncertainty. These uncertainties arise not only from ambipolar diffusion, but also from other diffusion processes. At higher altitudes (above 95 km), the anomalous diffusion dominates  (Ceplecha et al., 1998, Dyrud et al., 2001), while molecular diffusion dominates at lower altitudes (below 85 km) (Ceplecha et al., 1998, Lee et al., 2013). Hence, following Singer et al., (2008) and Ballinger et al., (2008), our study is confined to the altitude range between 85 and 95 km, as marked by bold horizontal dashed lines in figure 4.  From the figure, steeper negative slopes are observed above the turning altitude for both weak and strong  echoes. These negative slope of meteor decay time above the turning altitude near the mesopause region can be used in temperature estimation from the background temperature gradient height (Hocking, 1999), without considering the pressure information.  At Esrange, the negative slope is not so steeper in Northern summer as in case of other seasons. Also, at Rothera, somewhat steeper negative slopes are observed for strong echoes above the turning altitude, except in the Southern summer. Here, we can notice a clear seasonal variation in the decay time turning altitude,  which mainly depends on background mesospheric temperature, pressure and electron density. The temperature in the MLT-region significantly influences the plasma diffusion due to its direct relation with ambipolar diffusion coefficient (Ceplecha et al., 1998, Hocking, 1999). As a result, seasonal variation of temperature can lead to measurable changes in decay time turning altitude. Atmospheric pressure decreases exponentially with altitude, governed by the atmospheric scale height, which depends on temperature. Due to hemispheric differences in background temperature and seasonal dynamics, the rate of pressure decrease with altitude can vary between conjugate hemispheres, particularly in the MLT-region (Hocking, 1999). The decay time turning altitude is influenced not only by ambipolar diffusion but also by temperature dependent recombination processes. Specifically, molecular ion-electron recombination (90 - 110 km) and cluster ion-electron recombination (70 - 90 km) dominate at different altitude and temperature regimes, and contributing to rapid electron loss in meteor trails (Lee et al., 2013).  Seasonal variations in temperature modulate these recombination processes, thereby significantly affecting the altitude at which decay time transitions from being diffusion limited decay at high altitude to recombination limited decay at lower altitude. Consequently, this seasonal modulation of recombination plays a major role in shifting the decay time turning altitude.\\

For better visualisation of seasonal variation of decay time turning altitude, a box plot of seasonal median decay times prepared for both weak and strong echoes at Esrange and Rothera is shown in figure 5, which reveal a clear annual modulation consistent across hemispheres. Both sites exhibit longer decay times during winter and shorter ones in summer, reflecting reduced ambipolar diffusion and lower atmospheric densities in winter. Strong echoes persist longer than weak echoes throughout the year, indicating their higher electron line densities and reduced sensitivity to background turbulence. For strong echoes, asymmetric seasonal behaviour between the northern and southern stations suggests that mesospheric diffusion and recombination processes are predominantly governed by solar-driven variations in neutral temperature and density rather than by local geomagnetic influences.\\

Figure 4 and 5 also reveals notable hemispheric differences in seasonal variation of meteor decay time turning altitudes, particularly for strong echoes. These differences can be attributed to several factors, including variations in background ionospheric conditions, differences in magnetic field inclination, and differing levels of mesospheric gravity wave activity. Each of these processes influences the plasma diffusion and recombination, and thereby modulate the decay time turning altitude. The variations in background ionospheric conditions, specifically electron density, ion composition and ionospheric temperature structure play a key role in determining the meteor decay time turning altitude, separately for weak and strong echoes. Such variations between the opposite hemispheres contribute significantly to the observed asymmetry in decay time turning altitude (Kelly, 2009, Plane, 2012). Likewise, the variation of magnetic field inclination (dip angle) with latitudes and between the hemispheres plays a subtle yet important role in radar meteor observations and influencing the observed decay time turning altitude.  Although the Earth's magnetic field is broadly dipolar, the geomagnetic poles are not symmetrically aligned, resulting in hemispheric differences in magnetic field inclination and strength. In general, magnetic field lines tends to be steeper and in some regions, stronger in the Southern Hemisphere, enhancing the efficiency of ambipolar diffusion. This can contribute to faster meteor trail dissipation resulting in a lower decay time turning altitude compared to Northern Hemisphere (Hocking, 2004). Similarly, gravity wave activity also play a crucial role in modulating meteor decay time through its influence on mesospheric temperature, winds and turbulence. The topography and land-sea contrast are longer in Northern Hemisphere that leads to stronger gravity wave generation and more variability in mesosphere whereas its counter part experience less gravity wave drag (Fritts and Alexander, 2003), leading to colder and more stable mesospheric conditions result in systematically lower turning altitude.\\

In addition, several other factors may contribute to the seasonal and hemispheric variability in meteor decay time turning altitude. These include mesospheric wind dynamics, solar zenith angle and its associated radiative forcing, sudden stratospheric warming (SSW) events, particularly in the Northern Hemisphere and large-scale variations in sea surface temperature (SST). Each of these factors can modify the thermal and dynamical structure of the MLT-region, potentially impacting the balance between diffusion and recombination processes. However, these effects are beyond the scope of the present study and have not been explicitly considered in our analysis, but will be addressed in future work.

\section{Conclusions}

In this study, we have comparatively examined the seasonal variation of vertical distribution of meteor trail peak occurrence height and meteor decay time height, separately for weak and strong echoes recorded with two identical all-sky SKiYMET meteor radars located at conjugate polar latitudes,  Esrange (68$^\circ$N, Arctic) and Rothera (68$^\circ$S, Antarctica), during 2005 -  2019. Generally, the annual variation of meteor occurrence height agrees well with the earlier studies. Using the trail ionisation heights as a metric, we have recognised all the established major meteor showers of both the hemispheres, as the shower meteoroids likely to ionise at higher altitudes than the sporadic meteoroids. However, this method has some limitations and applicable only to shower the meteors having higher or lower ionisation heights than background sporadic meteors. \\

It has been observed that the vertical profiles of meteor occurrence height has a seasonal symmetric distribution at opposite polar latitudes, particularly for strong echoes, and there is about 1 km difference in occurrence height of weak and strong echoes. In contrary, the meteor decay time vertical profiles are showing an asymmetric distribution, whose altitude varies with seasons in the altitude range of 80 - 86 km. It is maximum in Northern winter and minimum in Northern summer at Esrange (68$^\circ$N), which is showing an exact opposite pattern at Rothera (68$^\circ$S).  By comparing, it is observed that there is no significant hemispheric difference in meteor occurrence height and meteor decay time, but only a seasonal dependent difference in decay time turning altitude. From the observations, it implied that the main physical processes that effect the meteor occurrence height and decay time turning altitude are similar at opposite latitudes (Younger et al., 2009). The maximum altitude during summer is consistent with colder temperatures that causes faster chemical reactions of electron removal, such as molecular ion - electron recombination, cluster ion - electron recombination, and electron attachment (Ballinger et al., 2008). All these episodes are consistent with the possibility of inter-hemispheric variability of  atmospheric density, pressure and temperature. In addition to ambipolar diffusion, the removal of meteor trail electrons by mesospheric ion chemistry plays an important role in the meteor trail decay process. Therefore, the seasonal changes in neutral density need to be considered before correctly computing the slope of decay time versus the height for temperature estimation (Kam et al., 2019).

%------------------------------------------------ TABLE 1 ------------------------------
\begin{table}
	\caption[solutions]{Identified meteor showers: Ordinal number since the beginning of the year, Name of the shower, Dates of the shower activity and height of the peak (Uq or Lq or both UqLq) detected at Esrange (68$^\circ$N) and Rothera (68$^\circ$S).   The unknown or newly identified showers are highlighted in bold.}
	\scriptsize
	\begin{tabular}{llllll}
		\hline 
		\hline\\
	\# &Shower & Esrange (68$^\circ$N) &Type & Rothera (68$^\circ$S) 		&Type\\\\
		\hline \hline \\
	1&	Quadrantids 				&	2-4 January 	&UqMLq	& ---	&--\\
	2&	\textbf{Unknown \#1}			&	---			& --		& \textbf{17-18 January} &Uq \\
	3&	\textbf{Unknown \#2}			&	\textbf{30-31 January}	&UqMLq	& ---	&--\\
	4&	$\gamma$-Ursae Minorids	&	15-25 January 	&MLq& ---			&--\\
	5&	Lyrids 					&	22-24 April 	&Uq& --- 	&--\\
	6&	$\eta$-Aquarids			&	--- 			&--	  & 4-6 May		&Uq\\
	7&	Piscis Austrinids			&	--- 			&--	  & 25-31 July	&Uq\\
	8&	Perseids 					&	12-15 August 	&UqMLq & --- 	&--\\
	9&	\textbf{Unknown \#3}			&	---			&UqMLq& \textbf{25-27 Sept.}	&Uq\\
	10&	Draconids					&	8-9 October	&MLq	  & ---			&--\\
	11&	Orionids 					&	17-27 October  	&UqMLq& ---	&--\\
	12&	Pheonicids 				&	---  			&--& 5-6 December	&Uq\\
	13&	$\sigma$-Hydrids 			&	11-12 December&Lq& ---	&--\\\\
		\hline
		\hline		
	\end{tabular}
\end{table}
%-------------------------------------------FIGURE 1 -----------------------------

\begin{figure*}
	\centering
{	
\includegraphics[height=10cm, width=8cm]{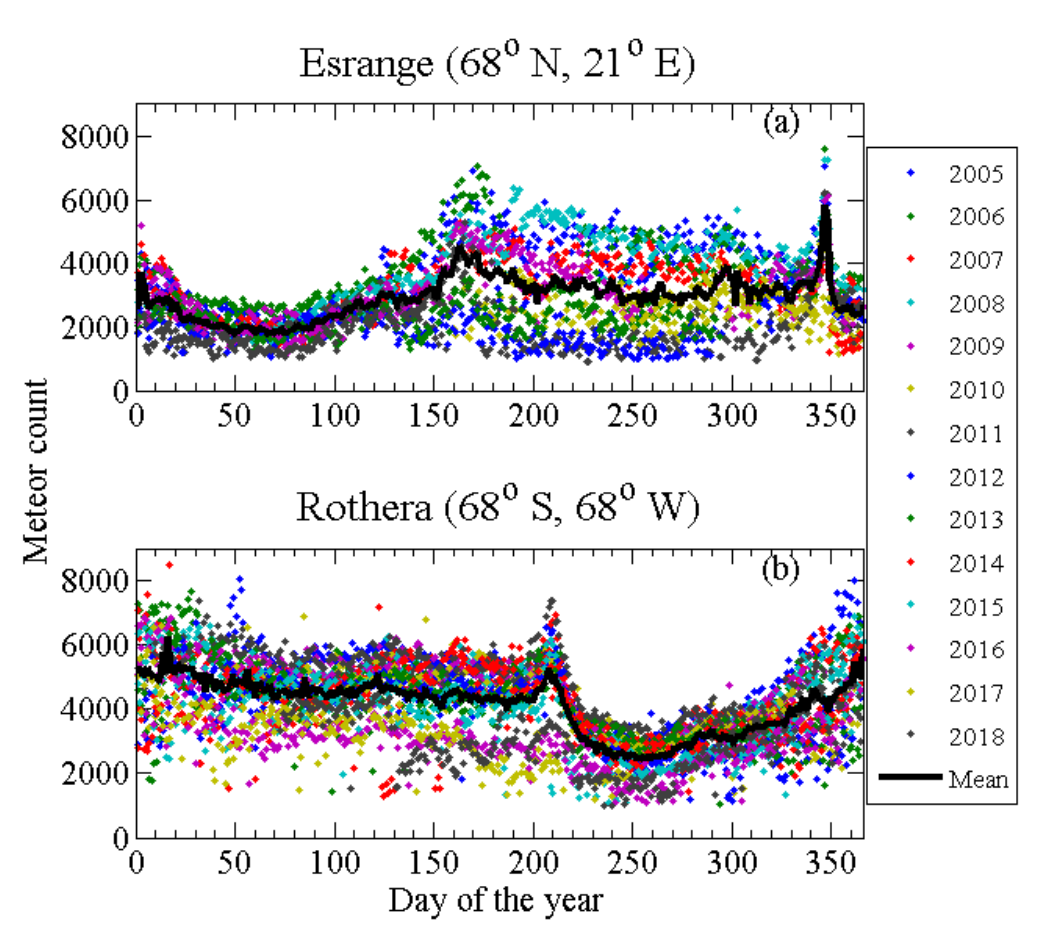}
}
\centering
{
\includegraphics[height=10cm, width=8cm]{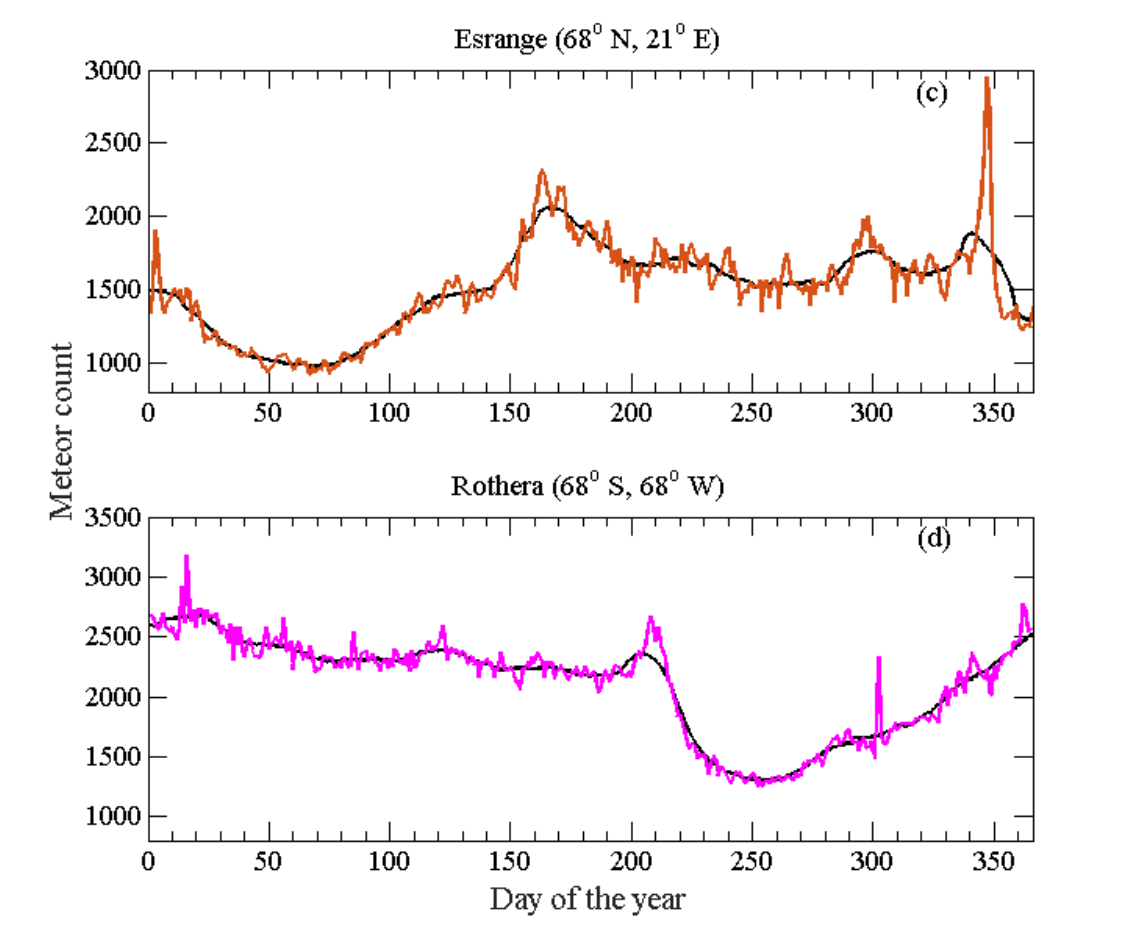}
}

\caption{The 15 year composite mean meteor count at Esrange (68$^\circ$N) and Rothera (68$^\circ$S) (i.e., average number of meteors detected per day) during 2005 - 2019. Black line in left side figures represent the yearly mean and in right side figure represents the 21-day moving mean in each case.}

\end{figure*}

%------------------------------------------End figure 1---------------------------
%----------------------------------------- FIGURE 2 ------------------------------

\begin{figure*}
\centering
{
\includegraphics[height=6cm, width=8cm]{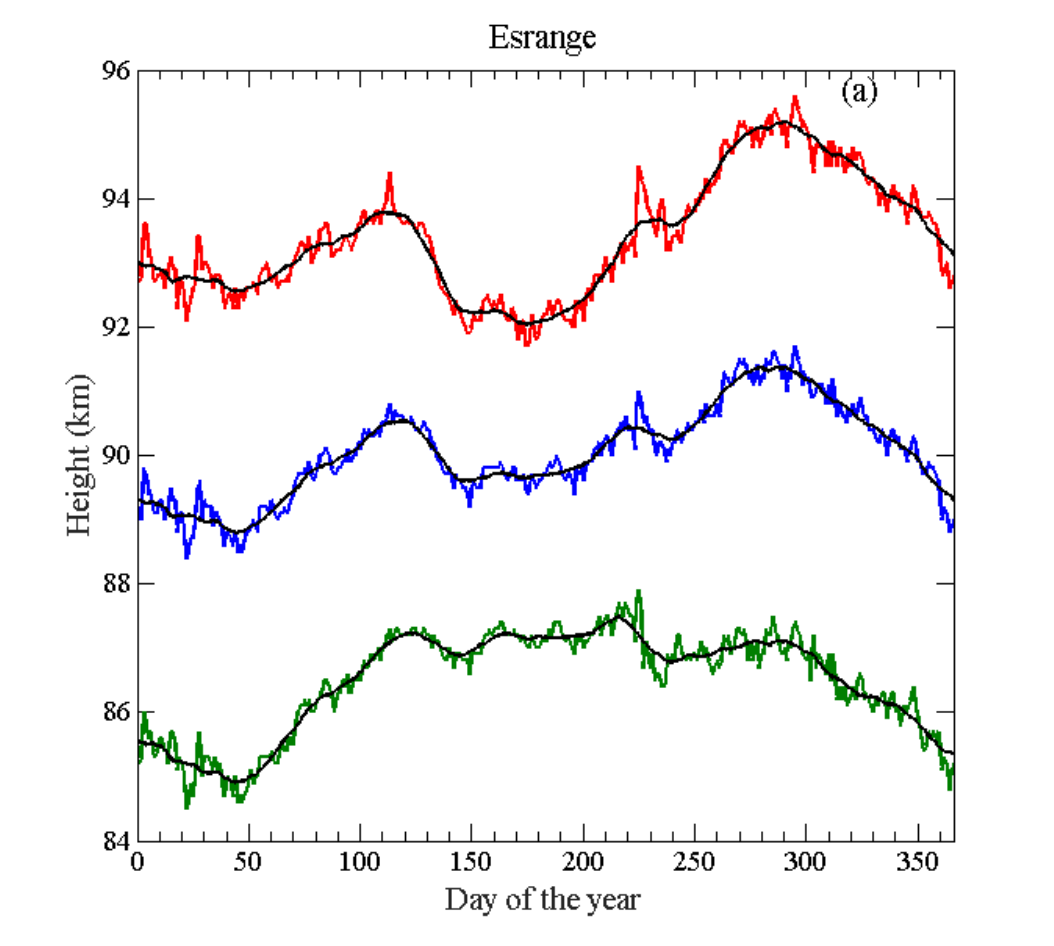}
}
\centering
{
\includegraphics[height=6cm, width=8cm]{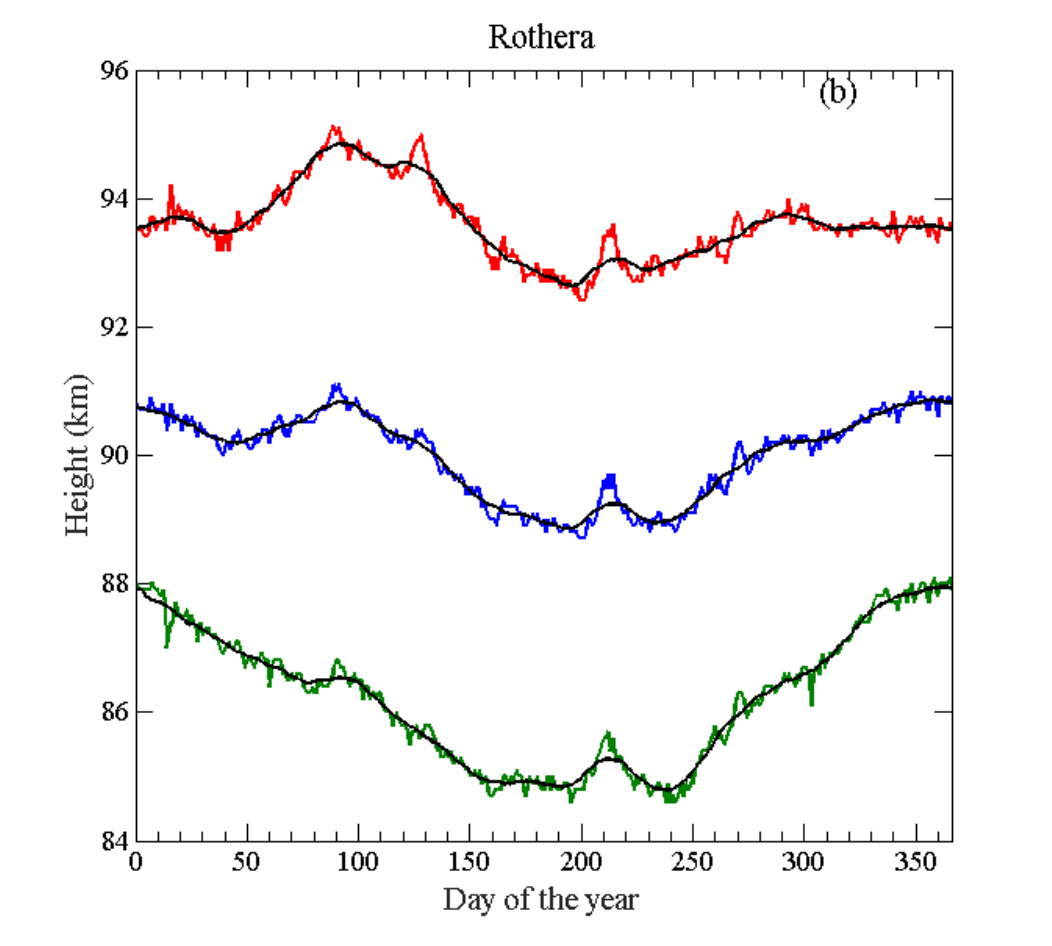}
}
\centering
{
\includegraphics[height=6cm, width=8cm]{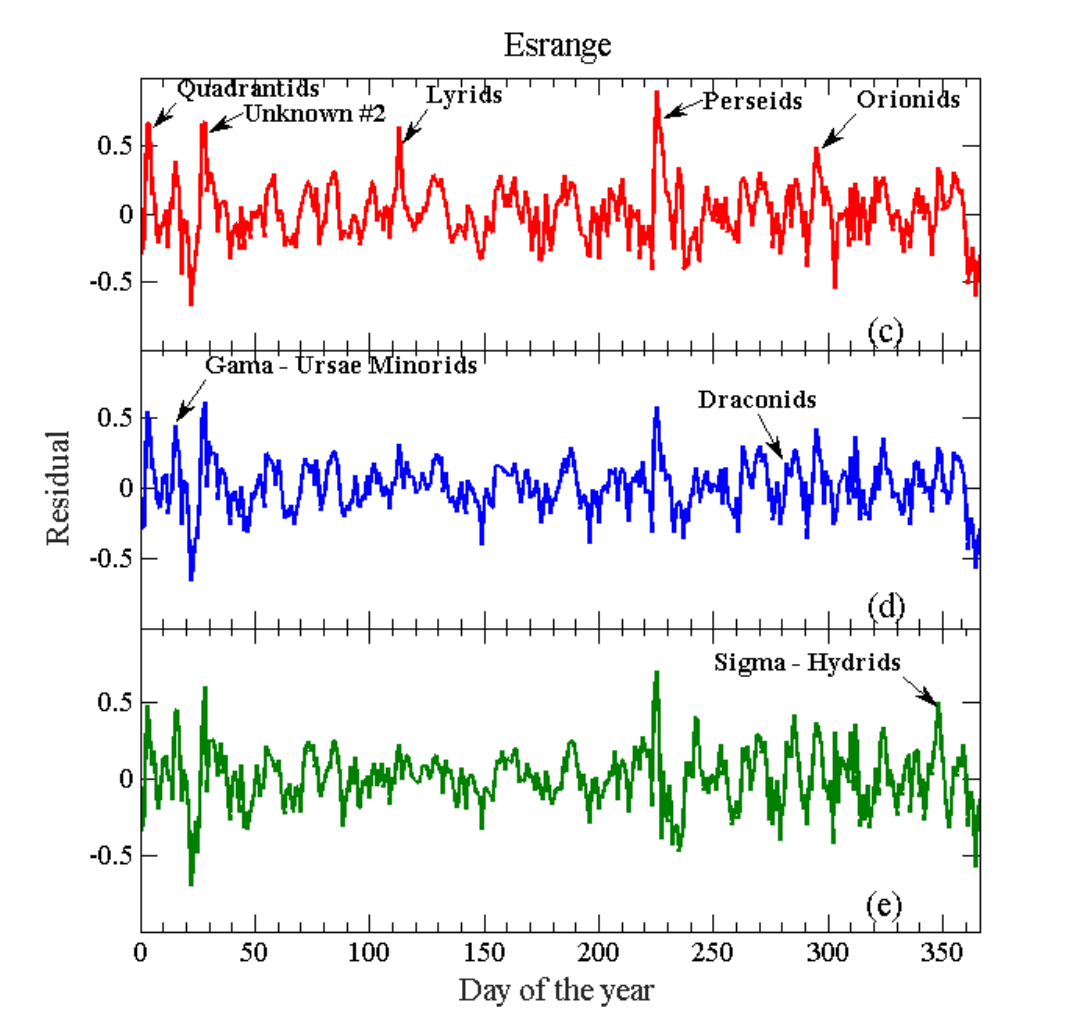}
}
\centering
{
\includegraphics[height=6cm, width=8cm]{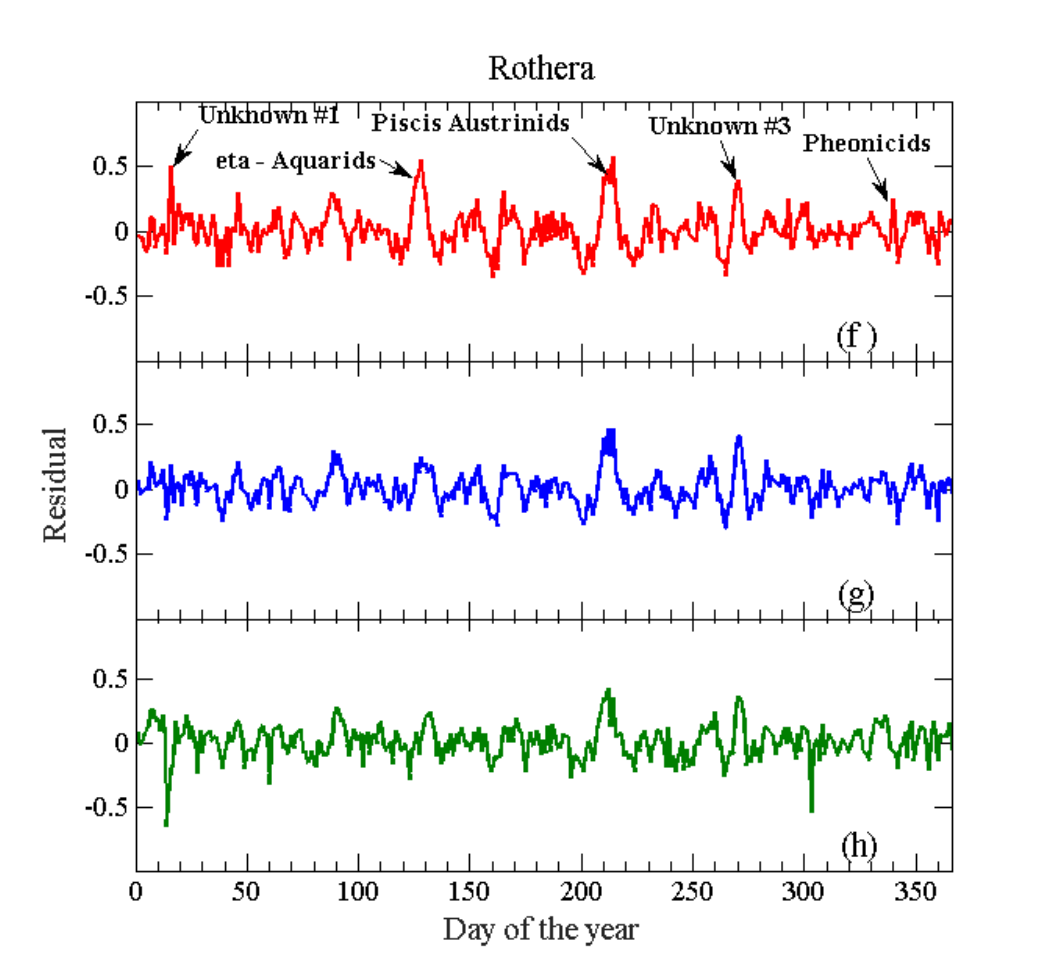}
}

\caption{The 15-year composite mean of the \textcolor{blue}{median (M)} and \textcolor{red}{upper quartile (Uq)} \& \textcolor{green} {lower quartile (Lq)} of meteor trail ionisation heights and their corresponding residual plots at Esrange (68$^\circ$N) and Rothera (68$^\circ$S) during 2005 -  2019. Black line in upper two panels represent the 21-day moving mean. The peaks in  corresponding residual plots are marked as major shower days.}

\end{figure*}
%------------------------------------------Figure 2---------------------------
%------------------------------------------Figure 3---------------------------

\begin{figure*}%[ht]
	\centering
	{
		\includegraphics[width=16.0cm, height=24.0cm]{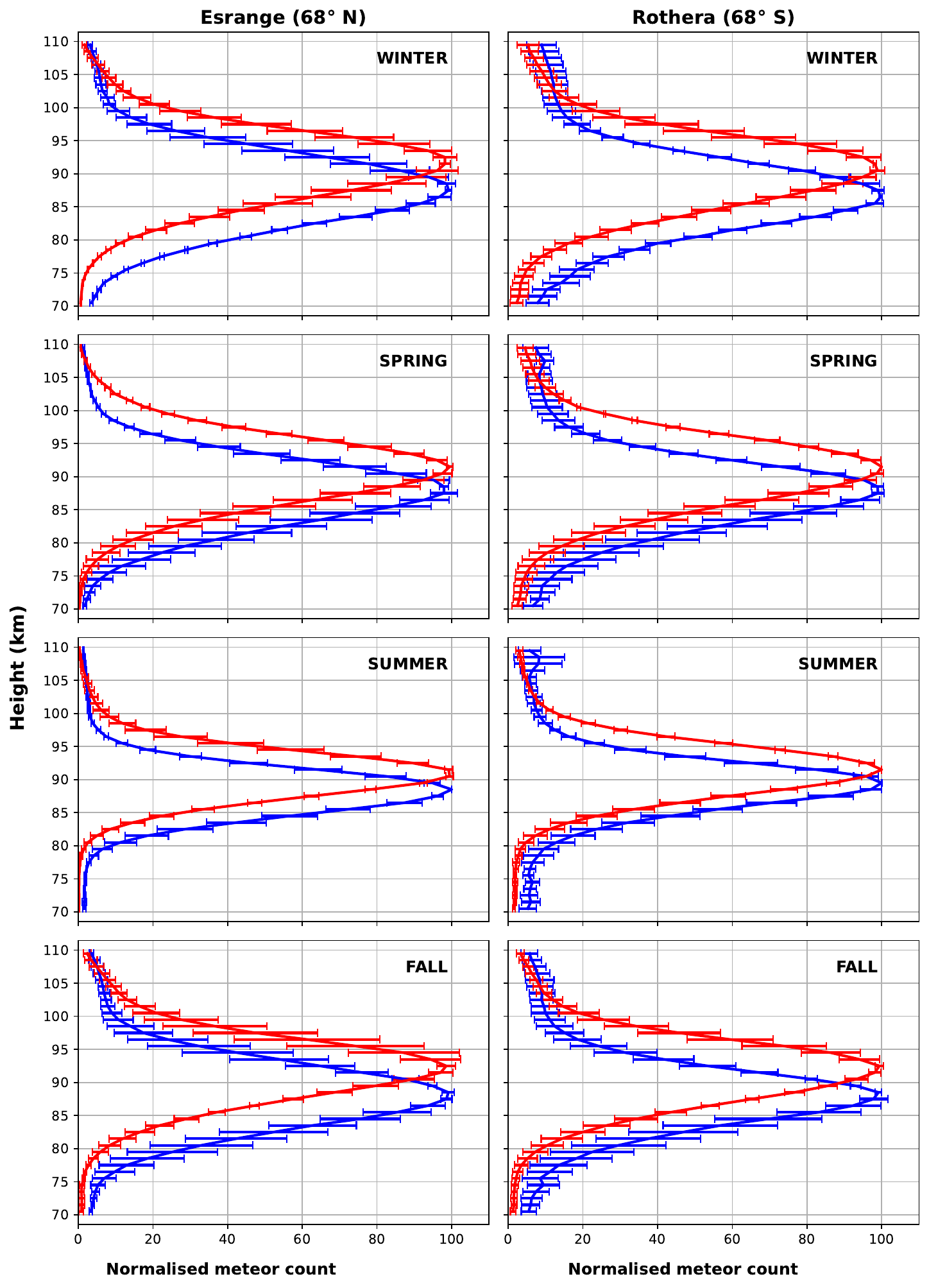}
	}

\caption{The 15-year composite mean of seasonal distribution of meteor occurrence height (km) for weak (\textcolor{blue}{--- blue line}) and strong (\textcolor{red}{--- red line}) echoes at  Esrange (68$^\circ$N) and Rothera (68$^\circ$S) during the period 2005 - 2019. Each data point represents the seasonal mean for each height bin and the error bars denote the corresponding standard deviation values.}

\end{figure*}

%------------------------------------------End figure 3----------------------
%------------------------------------------Figure 4---------------------------

\begin{figure*}%[ht]
	\centering
	{
		\includegraphics[width=16.0cm, height=24.0cm]{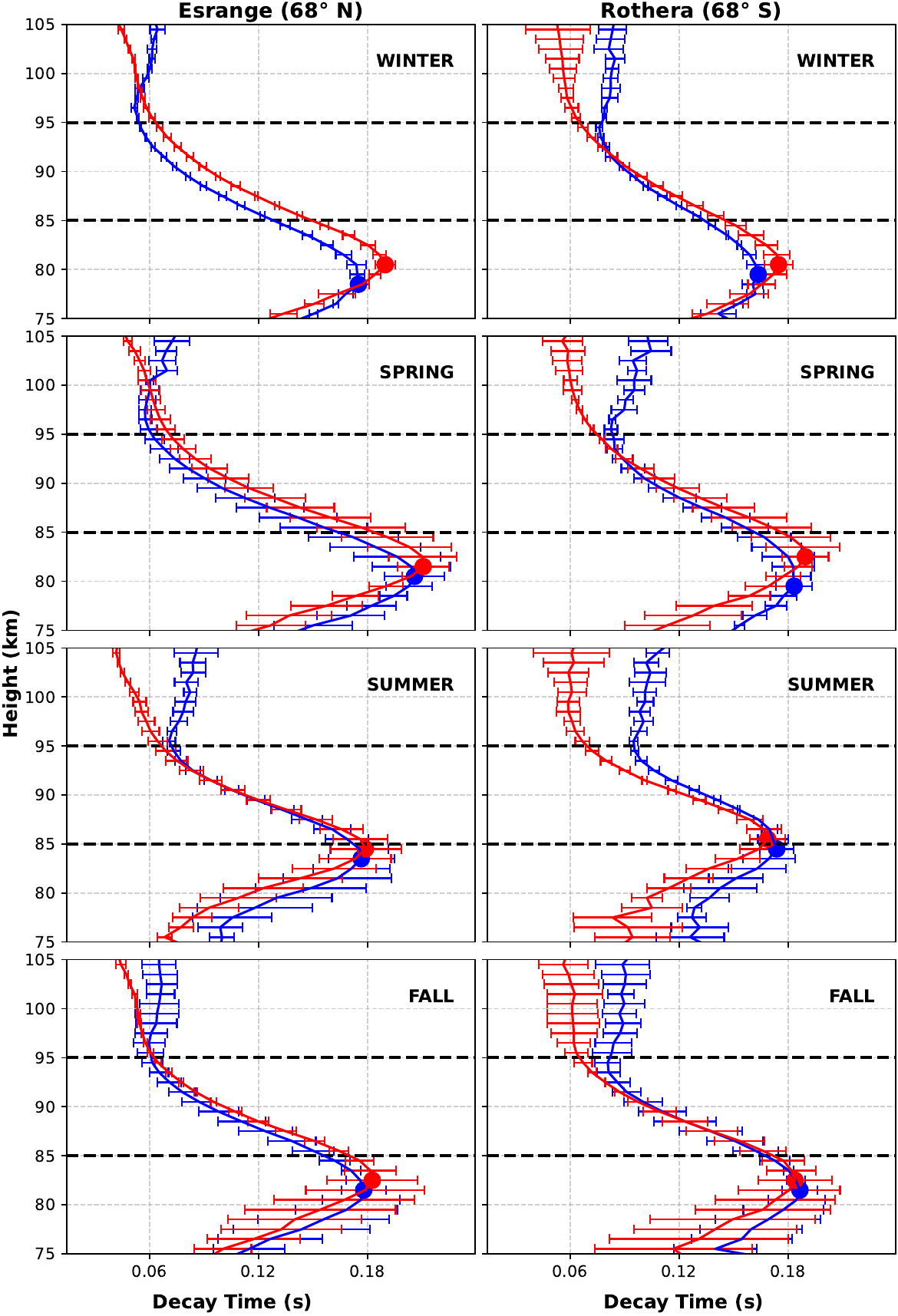}
	}
	
\caption{The vertical profiles of 15-year composite mean meteor decay time for weak (\textcolor{blue}{--- blue line}) and strong (\textcolor{red}{--- red line}) echoes at  Esrange (68$^\circ$N) and Rothera (68$^\circ$S) during the period 2005 - 2019. Each data point represents the seasonal mean of decay time and the error bars denote the corresponding standard deviation values. The data used in this study are confined to the altitude range between 85 and 95 km, as marked by bold horizontal dashed lines. For better visualisation, the turning altitudes in each case are marked with filled circles.}

\end{figure*}

%----------------------------------------End figure 4------------------------
%------------------------------------------Figure 5---------------------------

\begin{figure*}%[ht]
	\centering
	{
		\includegraphics[width=8.0cm, height=6.0cm]{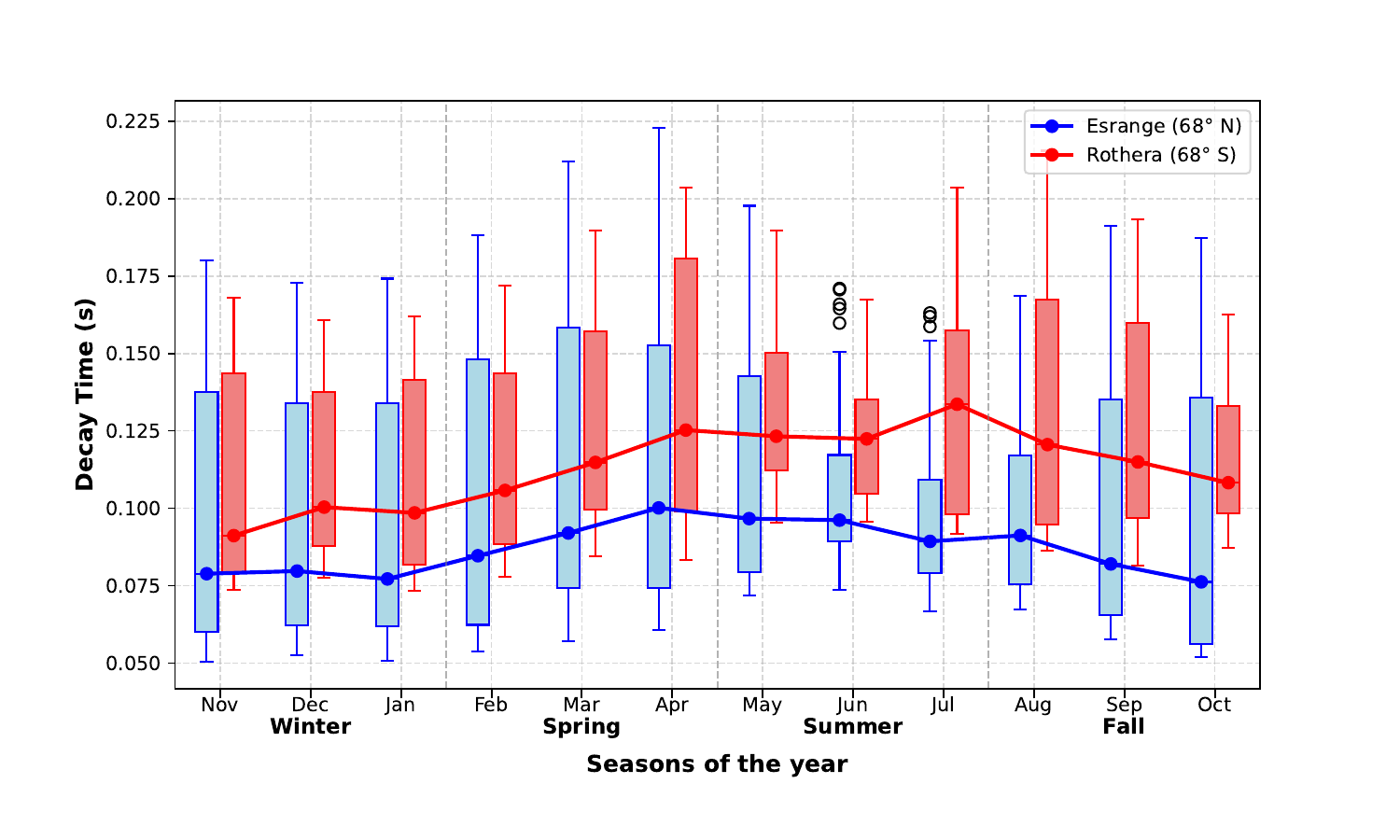}
	}
	\centering
	{
		\includegraphics[width=8.0cm, height=6.0cm]{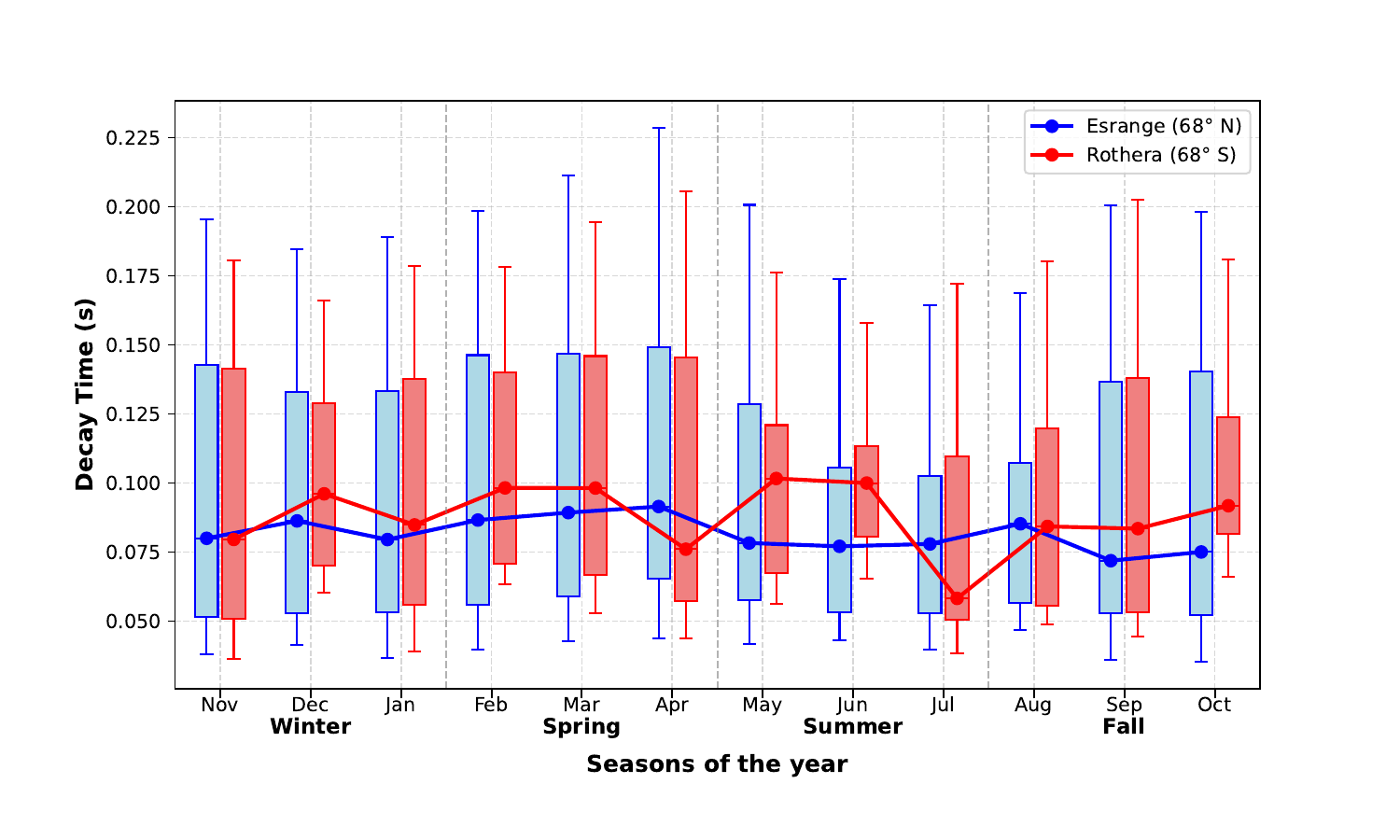}
	}
\caption{The seasonal representation of decay time for both weak (left figure) and strong (right figure) echoes at Esrange (68$^\circ$N) (\textcolor{blue}{--- blue line}) and Rothera (68$^\circ$S)(\textcolor{red}{--- red line}). The data from Rothera are presented with a shift of six months to allow an easy comparison of seasonal variation at opposite hemispheres.}

\end{figure*}
%-----------------------------------------End figure 5--------------------------------
\section*{Acknowledgement}
This work has used the archived data obtained from the Centre for Environmental Data Analysis Archive (CEDA) online service maintained by the British Antarctic Survey (BAS). I acknowledge the help of Prof N J Mitchell, University of Bath. I extend sincere gratitude to the anonymous referee for the constructive suggestions.

\section*{Data availability} Data used in this work can be accessed
through the University of Bath Skiymet meteor radar data collection maintained via BAS - CEDA (Mitchell, N.J., 2019). \cite{http://catalogue.ceda.ac.uk/uuid/836daab8d626442ea9b8d0474125a446}  \\\\\\\\\\\\\\\\\\\
%---------------------------------------End of Acknowledgement----------------

\end{document}